\documentstyle[psfig,aps,preprint]{revtex}
\newcommand{\th}{\Theta}
\newcommand{\pslash}{p\!\!\!/}

\begin{document}
\begin{titlepage}
\begin{flushright}
hep-ph/9703374{\hskip.5cm}\\
IOA-05-97
\end{flushright}
\begin{centering}
\vspace{.1in}
{\bf $b-\tau$ UNIFICATION AND NEUTRINO MASSES
IN $SU(5)$ EXTENSIONS OF THE MSSM WITH RADIATIVE
ELECTROWEAK BREAKING.} \\
\vspace{1 cm}
{\bf A. Dedes}
and {\bf K. Tamvakis}\\
\vspace{1 cm}
{\it Division of Theoretical Physics,
Physics Department,}\\
{\it University of Ioannina, GR-45110, Greece}\\ \vspace{0.5cm}

{\bf Abstract}\\
\end{centering}
\vspace{.1in}
We make a complete analysis of the Yukawa coupling unification
in $SU(5)$ extensions of the MSSM in the framework of the radiative
symmetry breaking scenario. Both logarithmic and finite threshold
corrections of sparticles have been included in
the determination of the gauge and Yukawa couplings at $M_Z$. The
effect of the heavy masses of each model in the renormalization
group equations is also included. We find that in the minimal
$SU(5)$ model $b-\tau$ Yukawa unification can be achieved
for too large a value of $\alpha_s$.
On the other hand the {\it
Peccei-Quinn} version of the {\it Missing Doublet} model,
with the effect of the right handed neutrino also included,
exhibits $b-\tau$ unification in excellent agreement with all
low energy experimental data. Unification of all Yukawa couplings
is also discussed.
\vspace{1.0cm}
\begin{flushleft}
March 1997 \\
\end{flushleft}
\vspace{0.05in}
\hrule
\vspace{0.05in}
E-mails\,:\,adedes@cc.uoi.gr,
tamvakis@cc.uoi.gr
\end{titlepage}
\section {Introduction}

Softly broken supersymmetry$\cite{Nilles}$, possibly resulting from an
underlying superstring framework, can lead to $SU(2)_L\times U(1)_Y$
gauge symmetry breaking through radiative corrections for a certain
range of values of the existing parameters$\cite{Iban}\cite{Ross}
\cite{castano}$. A most appealing feature of the Minimal Supersymmetric
Standard Model (MSSM) and its extensions is that it exhibits gauge coupling
unification$\cite{amaldi}$ in good agreement with the low energy values
of the three gauge couplings, known to the present experimental accuracy,
within the bounds set by the stability of the proton. The Higgs boson
running mass-squared matrix, although positive definite at very high
energies, when radiatively corrected develops at low energies a
negative mass-squared eigenvalue that triggers electroweak symmetry
breaking. Running parameters are studied in the framework of the
Renormalization Group equations.

In the present article we complete the analysis of electroweak
symmetry breaking in the MSSM and various $SU(5)$ extensions
of it$\cite{dedes1}$$\cite{dedes2}$, focusing on fermion masses.
Yukawa coupling relations, such as the $b-\tau$ Yukawa coupling
equality at high energies in relation to the experimental
$\frac{m_b}{m_{\tau}}$ value, give an extra constraint on models
in addition to existing strong constraints like the experimental
value of $\alpha_s(M_Z)$. The input parameters of our analysis, apart
from the standard low energy inputs ($\alpha_{EM}$, $G_F$, $M_Z$, quark
and lepton masses) are the soft breaking parameters ($M_o$, $M_{1/2}$,
$A_o$), the ratio of the two v.e.vs $tan\beta(M_Z)$, the sign of the
Higgs mixing parameter $\mu$ and the superheavy particle masses
($M_{H_c}$, $M_{\Sigma}$,...) entering through their thresholds in
the Renormalization Group equations.  On the other hand,
our output includes the strong
coupling $\alpha_s(M_Z)$, the complete sparticle spectrum as well as
other quantities, like the unification scale $M_{GUT}$, etc. In this
analysis, for obvious reasons, we shall concentrate on that part of
parameter space that leads to $\alpha_s$ compatible with experiment.
The experimental values of third generation fermion masses will
be introduced as input while
the high energy values of Yukawa couplings will be treated as an output.

We have included both
logarithmic and finite one loop corrections (see Ref.$\cite{dedes2}$
for more details) in the calculation of $sin^2\theta(M_Z)|_{\overline{DR}}$
that determines
the $\hat{g}_1|_{\overline{DR}}$ and $\hat{g}_2|_{\overline{DR}}$
gauge coupling values which are taken as boundary
conditions at $M_Z$. Yukawa couplings are determined
with the same next to leading order accuracy as we employed
 in our analysis for the
gauge couplings. We extract the $\overline{DR}$ values of the Yukawa couplings
at $M_Z$
 from the corresponding pole masses of quarks and leptons
$\cite{Bagger}$ as we describe below.

The fermion masses, defined as the poles of the
corresponding propagators, are related to the $\overline{DR}$ masses,
$\hat{m}_f(Q)$, by the self energies, $\Sigma_f(\pslash)$, as
\begin{equation}
m_f^{pole}=\hat{m}_f(Q)+\Re e\Sigma_f(m_f^{pole})
\end{equation}
where $-i\Sigma_f(m_f^{pole})=-i(\Sigma_1+m_f^{pole} \Sigma_{\gamma})$
is
the one loop self energy on shell of quarks or leptons.
Our results for self energies agree with those of references
$\cite{Bagger2}$$\cite{donini}$.
In the case of
the {\it Peccei-Quinn} version of the {\it Missing Doublet Model} (MDM+PQ)
the presence of the right handed neutrino of mass $M_R$ gives
rise to a mass for the left handed neutrino
$m_{\nu}=\frac{Y_{{\nu}_{\tau}}^2 v^2 sin\beta}{M_R}$ via the see-saw
mechanism $\cite{Slansky}$. The calculated
one loop neutrino self energy correction is denoted by
$\Sigma_{\nu}$.
{}From equation (1) we arrive at the values of Yukawa couplings
at $M_Z$
\begin{equation}
\hat{Y}_b(M_Z)|_{\overline{DR}}=\frac{m_b^{pole}-
\Re e\Sigma_b(m_b^{pole})|_{\overline{DR}}}{\hat{v}(M_Z)|_{\overline{DR}}
cos\beta(M_Z)}
\end{equation}
\begin{equation}
\hat{Y}_{\tau}(M_Z)|_{\overline{DR}}=
\frac{m_{\tau}^{pole}-
\Re e\Sigma_{\tau}(m_{\tau}^{pole})
|_{\overline{DR}}}{\hat{v}(M_Z)|_{\overline{DR}}
cos\beta(M_Z)}
\end{equation}
\begin{equation}
\hat{Y}_t(M_Z)|_{\overline{DR}}=\frac{m_t^{pole}-
\Re e\Sigma_t(m_t^{pole})|_{\overline{DR}}}{\hat{v}(M_Z)|_{\overline{DR}}
sin\beta(M_Z)}
\end{equation}
\begin{equation}
\hat{Y}_{\nu}(M_Z)|_{\overline{DR}}=\sqrt{\frac{M_R (m_{{\nu}_{\tau}}^{pole}
-\Re e \Sigma_{{\nu}_{\tau}}(m_{{\nu}_{\tau}}^{pole})|_{\overline{DR}})}
{\hat{v}^2(M_Z)|_{\overline{DR}}
 sin^2\beta}}
\end{equation}
where all hats mean running couplings or masses evaluated in the
$\overline{DR}$ scheme. The running vacuum expectation value is given
by the following formula,
\begin{equation}
M_Z^2+\Re e\Pi_{ZZ}^{T}(M_Z^2)=\frac{1}{2} (\hat{g}_1^2(M_Z)+
\hat{g}_2^2(M_Z))
\hat{v}^2(M_Z)
\end{equation}
where $M_Z^2$ refers to the
experimental Z-boson mass and $\Pi_{ZZ}^T$ is the real transverse part
of the Z-boson one loop corrections at $M_Z$.

For a given set of pole masses $m_t^{pole}$, $m_b^{pole}$, $m_{\tau}^{pole}$
we define the $\overline{DR}$
Yukawa couplings at $M_Z$. Then we use the 2-loop Renormalization Group
equations to run up to the scale $M_{GUT}$, where $\hat{g}_1$ and
$\hat{g}_2$ meet. We impose the unification
condition
\begin{equation}
g_{GUT}=\hat{g}_1(M_{GUT})=\hat{g}_2(M_{GUT})=\hat{g}_3
(M_{GUT})
\end{equation}
and run down to $M_Z$. The thresholds of the heavy masses, both in
gauge and Yukawa couplings, are treated with by using the well known method
 of the step function approximation $\cite{lah}$$\cite{dedes1}$.
We have assumed universal boundary conditions for
the soft breaking parameters $A_o$, $M_o$, $M_{1/2}$. The whole
procedure is iterated until convergence is reached, imposing
the constraints of radiative symmetry breaking, the proton decay bound,
the experimental bounds on supersymmetric particles and the
perturbativity of the Yukawa and gauge couplings up to $M_P$.

The value of the bottom quark pole
mass can be determined indirectly through its influence on hadron
properties. In Table I we display the pole mass of the bottom quark
as it is
extracted from experiment in different processes following various
techniques
$\cite{booklet}$.
(For the determination of the pole mass of the bottom quark from QCD
momentum sum rules for the $\Upsilon$ system see the recent article of
Ref.$\cite{pich}$.)
\begin{center}
\begin{tabular}{cc}\hline
\multicolumn{2}{c}{\bf TABLE I} \\ \hline
{\it Description} & $m_b (GeV)$\\
Lattice computation of $\Upsilon$ spectroscopy   &$5\pm 0.2$ \\
$e^+e^-\rightarrow b$ hadrons cross sections &$4.827\pm0.007$ \\
Heavy quark effective theory	&$4.61\pm 0.05$ \\
$\Upsilon$ system		&$4.60\pm 0.02$ \\
\hline \hline
\end{tabular}
\end{center}
\centerline{\footnotesize {\bf Table I:} Values of the bottom quark pole mass}
\vspace{1cm}

As a characteristic mean value of the bottom quark mass
we shall choose below
$m_b=4.9 GeV$.
The experimental mass of $\tau$-lepton is
$m_{\tau}=1.777 GeV$. In addition, we adopt for the mass of the top quark
the average experimental value of $D\emptyset$
 and $CDF$ experiment $m_t=180 GeV$.
We also take the $\tau$-neutrino to be in the cosmologically
interesting domain $m_{{\nu}_{\tau}}=3-10 eV$. When we depart from
the above values we indicate so.

The unification of the Yukawa couplings in the MSSM or in some
supersymmetric Grand Unified Models has been studied in
$\cite{Bagger}$$\cite{carena}$$\cite{wright}$.
The presence of  right-handed
neutrinos as the only thresholds in the grand desert is
studied in $\cite{visani}$ while some of
their consequences in a supersymmetric $SO(10)$ model in $\cite{brig}$.
In this article we examine, in a higher level accuracy
framework, the ratios of the
Yukawa couplings at the Unification scale $M_{GUT}$ in the
minimal $SU(5)$ and the Peccei-Quinn version of
the Missing Doublet Model, in which the right-handed neutrino is
included.

\section {A little bit more about the minimal SU(5) model}

In this section we apply the preceeding formalism
on the minimal supersymmetric $SU(5)$ model.
We start with the superpotential $\cite{dimopoulos}$,
\begin{eqnarray}
\cal W &=& \frac {1}{2} M_1 Tr(\Sigma^2)+\frac{1}{3}\lambda_1
Tr(\Sigma^3)+M_2\overline{H} H+\lambda_2 \overline{H}\Sigma H \nonumber
\\[1.5mm]&+&\sqrt{2}
Y_{(d)}^{ij}\Psi_i\phi_j\overline{H}-\frac{1}{4}Y_{(u)}^{ij}\Psi_i\Psi_j
H \;\;\;\;\;\; i,j=1..3
\end{eqnarray}
where the superfields $\Sigma$, $H$, $\overline{H}$, $\phi$, $\Psi$
belong to the representations {\bf 24}, {\bf 5}, $\overline{\bf 5}$,
$\overline{\bf 5}$,
and {\bf 10} respectively. The $SU(5)$ breaking occurs when the $\Sigma$
superfield develops a vacuum expectation value in the direction
$<\Sigma>\equiv V\,Diag(2,2,2,-3,-3)$. In the resulting
 effective $SU(3)\times SU(2)
\times U(1)$ theory  we get
the following heavy particles with quantum numbers:
\begin{equation}
M_{H_c}^{(3,1,-\frac {1}{3})}=M_{\overline{H}_c}^{(3,1,\frac {1}{3})}=5
\lambda_2 V
\end{equation}
\begin{equation}
M_{\Omega}^{(8,1,0)}=M_{\omega}^{(1,3,0)}\equiv M_{\Sigma}=5 \lambda_1 V
\quad , \quad M_S^{(1,1,0)}=
\frac{M_{\Sigma}}{5}
\end{equation}
\begin{equation}
M_V=5 g_5 V
\end{equation}
In addition to the RGE's for the gauge
couplings $\cite{dedes2}$, we display here the RGE's for the Yukawa
couplings, treating the superheavy thresholds in the step function
approximation
\begin{eqnarray}
\hspace{0.3cm}16 \pi^2 \frac{dY_b}{dt}& = &Y_b \{6 Y_b^2 + Y_t^2 + Y_{\tau}^2
+ 3 Y_b^2 \th(\overline{H}_c) + 2 Y_t^2 \th(H_c) + \frac{3}{2} \lambda_2^2
\th(\omega) \nonumber \\[3mm]&+&\frac{3}{10} \lambda_2^2 \th(S) +
3 \lambda_2^2 \th(V) -
( \frac{7}{15} g_1^2 + 3 g_2^2 + \frac{16}{3} g_3^2) - 8 g_5^2 \th(V)\}
\\[10mm]
\hspace{0.3cm} 16 \pi^2 \frac{dY_{\tau}}{dt} & = & Y_{\tau}
\{ 4 Y_{\tau}^2
+ 3 Y_b^2 +\frac{3}{2} \lambda_2^2 \th(\omega) + \frac{3}{10} \lambda_2^2
\th(S) + 3 \lambda_2^2 \th(V) \nonumber \\[3mm] &+& 3 Y_b^2
\th(\overline{H}_c) +
3 Y_t^2 \th(\overline{H}_c)-(\frac{9}{5} g_1^2 + 3 g_2^2 )- 12 g_5^2 \th(V)
\}
\\[10mm]
\hspace{0.3cm} 16 \pi^2 \frac{dY_t}{dt} & = & Y_t \{ 6 Y_t^2 +
Y_b^2 + 3 Y_b^2 \th(\overline{H}_c) + 3 Y_t^2 \th(H_c) +  \frac{3}{10}
\lambda_2^2 \th(S) \nonumber \\[3mm] &+& \frac{3}{2} \lambda_2^2 \th(\omega) +
3 \lambda_2^2
\th(V) -(\frac{13}{15} g_1^2 + 3 g_2^2 + \frac{16}{3} g_3^2) -
10 g_5^2 \th(V) \} \\[10mm]
\hspace{0.3cm} 16 \pi^2 \frac{d\lambda_1}{dt} & = & 3 \lambda_1
\th(\omega) \{ \frac{6}{5} \lambda_1^2 \th(S) +
3 \lambda_1^2 \th(V) +
\lambda_2^2 - 4 g_2^2 -6 g_5^2 \th(V)\}\\[10mm]
\hspace{0.3cm} 16 \pi^2 \frac{d\lambda_2}{dt} & = & \lambda_2
\th(\omega) \{3 Y_b^2 + Y_{\tau}^2 + 3 Y_t^2 +3 \lambda_2^2 \th(\omega) +
\frac{3}{5} \lambda_2^2 \th(S)+ 6 \lambda_2^2 \th(V) \nonumber \\[3mm] &+&
\frac{6}{5}
\lambda_1^2 \th(S) + 3 \lambda_1^2 \th(V) + \lambda_2^2 - (\frac{3}{5}
g_1^2 + 7 g_2^2 ) -12 g_5^2 \th(V) \}
\end{eqnarray}
where $t=ln\frac{Q}{M_{GUT}}$. We have denoted $\th(x)\equiv \th(Q^2-
M_x^2)$.
Note that $M_{GUT}=max\{M_V,M_{H_c},M_{\Sigma}\}$ which we take
equal to $M_V$ as it turns out to be in all relevant cases $\cite{dedes2}$.

In Figure 1 we have plotted the ratio of bottom and tau Yukawa couplings
at $M_{GUT}$ as a function of $tan\beta$ and the input high energy mass
$M_{\Sigma}$.
We select as input soft breaking masses at $M_{GUT}$, so as to
give consistent low energy results within the experimentally
acceptable region.
We obtain that the  $b-\tau$ unification in the minimal SU(5) is in
agreement with the pole mass of the bottom quark extracted from the Lattice
computation of $\Upsilon$
 spectroscopy as it is shown in Table I. Nevertheless, even
in
this case the output $\overline{MS}$ value of $\alpha_s(M_Z)$ is
quite large ($\alpha_s(M_Z)>0.127$) compared with its
average experimental value $0.118\pm 0.003$
$\cite{booklet}$.
Both the proton decay bound and the theoretical
perturbativity bound on the top Yukawa coupling ($Y_t<1.5$ at
$M_{GUT}$) strongly constrain the Yukawa unification. We have also
seen that Yukawa unification is sensitive through $\alpha_s$ to
the logarithmic and finite self-energy corrections of Yukawa
couplings as well as to the analogous
corrections to
the Weinberg angle at $M_Z$. As it is obvious
 from Figure 1, decreasing the
value of
$M_{\Sigma}$ destroys (for the lower values of $m_b$) or restores (for higher
values of $m_b$) the picture of Yukawa unification in contrast to
the MSSM case, in which all the high
energy particles have been decoupled from the renormalization
group equations. In Ref.$\cite{bachas}$ the hope was expressed that
lowering the value of $M_{\Sigma}$ might result in the
unification of gauge couplings at a scale close to
the so-called string scale $M_{string}=5\times10^{17} GeV$. This
 can be the case only if $m_b\geq 5 GeV$, as shown
in Figure 1, but again with too large
a value for $\alpha_s$ $\cite{dedes2}$.
 A proposed mechanism to reconcile the value of $\alpha_s$
with the experimental data is the one which
goes through the high energy threshold of colour triplet mass $M_{H_c}$.
Lowering this value down to $10^{15} GeV$ we obtain values of $\alpha_s(M_Z)$
which are in good agreement with the experimental ones.
This cannot be the case however in
minimal SU(5) where we  encounter
the bound $M_{H_c}\geq 1.8\times 10^{16}$ GeV
due to the proton decay constraint. This problem is solved
in the {\it Peccei-Quinn} version of
the {\it Missing Doublet Model} where the
proton decay rate is suppressed due to
the presence of the intermediate
scale $10^{12} GeV$ .
Putting aside
the problem of
large $\alpha_s$ in the minimal $SU(5)$ model
and varying $M_{H_c}$ in the region $(1.8-3)\times 10^{16} GeV$
we obtain values for the $b-\tau$ Yukawa ratio,
inside the shaded region of Figure 1.

\section {The  Peccei-Quinn version of the  Missing
Doublet model}

In order to avoid the numerical fine-tuning required in the
minimal $SU(5)$ model we can replace the adjoint Higgs
representation by the {\bf 75} representation
which couples the Higgs pentaplets
to an extra pair of Higgses $\Theta$ and
$\overline{\Theta}$ in the
{\bf 50}+ $\overline{\bf {50}}$ representation respectively
that contains no isodoublets.
 This extension
of the minimal $SU(5)$ is known as the {\it Missing Doublet}
model $\cite{masiero}$. The superpotential of this model
is
\begin{eqnarray}
\cal W &=& M_{1} Tr(\Sigma^2)+\frac{1}{3} \lambda_{1} Tr(\Sigma^3)+\lambda_2 H
\Sigma\Theta+\overline{\lambda}_2\overline{H}\Sigma\overline{\Theta}+M_{2}
\overline{\Theta}\Theta \nonumber \\[1.5mm] &-& \frac{1}{4} Y_{(u)}^{ij}
\Psi_{i}\Psi_{j}H+\sqrt{2} Y_{(d)}^{ij}\Psi_{i}\phi_{j}\overline{H} \;\;\;\;\;
\; i,j=1..3
\end{eqnarray}
As it has been shown in Ref.$\cite{dedes2}$ this model built in
such a  way, although it predicts small values of $\alpha_s(M_Z)$, leaves
a narrow window in the parametric space (especially in the
$tan\beta$) to work with, because of the bound which
comes from the proton decay stability. This problem which is also
present in the minimal $SU(5)$, albeit
 with large values of $\alpha_s(M_Z)$,
provided strong motivation to construct versions of $SU(5)$ with
a {\it Peccei-Quinn} symmetry $\cite{peccei}$ that naturally
suppresses $D=5$ operators by a factor proportional to the ratio
of the {\it Peccei-Quinn} breaking scale over the GUT scale $\cite{hisano}$.
If we promote the symmetry
by an extra global {\it Peccei-Quinn}
$U(1)$ factor the terms responsible for Higgs masses must be
replaced by
\begin{eqnarray}
\lambda_2 H \Sigma \Theta &+& \overline{\lambda}_2 \overline{H}
\Sigma \overline{\Theta} + \lambda^{\prime}_2 H^{\prime} \Sigma
\Theta^{\prime}+\overline{\lambda}^{\prime}_2 \overline{H}^{\prime}
\Sigma \overline{\Theta}^{\prime} \nonumber \\[2mm] \nonumber
&+& M_2 \Theta \overline{\Theta}^{\prime}+ M_2^{\prime} \Theta^{\prime}
\overline {\Theta} + \lambda_3 P \overline{H}^{\prime} H^{\prime}
\end{eqnarray}
We have introduced a new set of chiral superfields,$H'({\bf 50})$,
$\overline{H}'({\bf \overline{50}})$, $\Theta'({\bf 50})$,
$\overline{\Theta}'({\bf \overline{50}})$ and $P$ is a gauge singlet
superfield. The charges under the {\it Peccei-Quinn symmetry} are
$\Psi (1)$, $\phi (-1/2)$, $H(-2)$,
$\overline{H}(-1/2)$,
$\Theta(2)$, $\overline{\Theta}(1/2)$,
$\Theta^{\prime}(-2)$, $\overline{\Theta}^{\prime}(-1/2)$,
$\overline{H}^{\prime}(1/2)$,
$H^{\prime}(2)$, $P(-5)$.
The {\it Peccei-Quinn} symmetry is broken
{\footnote {This is achieved through a
suitable sector involving singlets
$\cite{murayama}$.}} when the gauge singlet $P$
develops a vacuum expectation value at an intermediate energy $<P>\equiv
\frac{M_{H_f'}}{\lambda_3}\sim 10^{10}-10^{12} GeV$.
The model is naturally extended
by introducing the three right-handed neutrino superfields
$N_i^c$ adding to the superpotential the following Yukawa interactions:
\begin{equation}
{\cal{W}}_N=Y_{\nu} N^c \phi H + \frac{1}{2} \lambda_4 N^c N^c P
\end{equation}
These terms induce Majorana masses for the right-handed neutrino
multiplets $N_i^c$ and these masses are proportional to the
intermediate breaking scale since $M_R=\frac{\lambda_4}{\lambda_3}
M_{H_f'}\sim O(10^{11}-10^{12}) GeV$.
 Finally, the terms incorporated in
${\cal{W}}_N$ induce through the see-saw mechanism $\cite{Slansky}$
very small masses for the neutrinos.
The spectrum obtained after the breaking of the $SU(5)$ gauge group
can be read from Table II
where we have used the following reasonable assumptions
 for the parameters
in the superpotential of this model:
\begin{equation}
{\lambda}_2 \simeq \overline{\lambda}_2 \simeq {\lambda}_2'
\simeq \overline{\lambda}_2' \hspace{0.8cm} M_2\simeq M_2'\sim 10^{18} GeV
\end{equation}
\begin{center}
\begin{tabular}{|c|c|c|} \hline
\multicolumn{3}{|c|}{\bf TABLE II} \\ \hline
75  &   50  &  Colour Triplets \\[2mm] \hline
$M_{S_8}(8,3,0)\equiv M_{\Sigma}$ &$ M(6,1,\frac{4}{3})=4 M_2$ &
$M_{H_c}\simeq M_{H_c'}\simeq 32 \lambda_2^2 \frac{V^2}{M_2}$ \\ [2mm]\hline
$M_S(1,1,0)=0.4 M_\Sigma $& $M(8,2,\frac{1}{2})=8 M_2$ & $M_{\th_c}
\simeq M_{\th_c'}\simeq 8 M_2$
\\[2mm] \hline
$M_{S_6}(6,2,\frac{5}{6})
=M_{\overline{S}_6}(\overline{6},2,-\frac{5}{6})=
0.4 M_\Sigma$& $M(\overline{6},3,-\frac{1}{3})= 4 M_2 $& Doublets
\\[2mm] \hline
$M_{\cal{O}}(8,1,0)=0.2 M_\Sigma$ & $M(\overline{3},2,-\frac{7}{6})
=8 M_2$ & $M_{H_f}=0$ \\[2mm] \hline
$M_{S_3}(3,1,\frac{5}{3})=
M_{\overline{S}_3}(\overline{3},1,-\frac{5}{3})=
0.8 M_\Sigma$ & $M(1,1,-2)= 8 M_2$ & $M_{H_f'}=
\lambda_3 <P>\simeq 10^{11} GeV$ \\[2mm] \hline
$M_\Sigma \equiv 20 M_1=-\frac{40}{3} \lambda_1 V$ &   &
 $M_V=\sqrt{6} g_5 V$
\\ [2mm]\hline
\end{tabular}
\end{center}
\centerline{\footnotesize {\bf Table II:} The spectrum of the MDM+PQ model}
\vspace{1cm}

The running of the Yukawa couplings at the one loop level is given
by the following renormalization group equations,
\begin{eqnarray}
16 \pi^2 \frac{dY_b}{dt}& = &Y_b \{6 Y_b^2 + Y_t^2 +
Y_{\tau}^2+ \frac{3}{2} Y_b^2 \th(H_c) + \frac{3}{2} Y_b^2 \th(H_c') +
Y_t^2 \th(H_c) \nonumber \\[3mm]&+&Y_t^2 \th(H_c')+Y_{\nu}^2 \th(H_c)
\th(R)
-( \frac{7}{15} g_1^2 + 3 g_2^2 + \frac{16}{3} g_3^2) - 8 g_5^2 \th(V)\}
\\[10mm]
16 \pi^2 \frac{dY_{\tau}}{dt} & = & Y_{\tau}
\{ 4 Y_{\tau}^2 + 3 Y_b^2  +\frac{3}{2} Y_b^2
\th(H_c) +\frac{3}{2} Y_b^2 \th(H_c')+\frac{3}{2} Y_t^2 \th(H_c)
\nonumber\\[3mm] &+&
\frac{3}{2} Y_t^2 \th(H_c')+Y_{\nu}^2 \th(R)
-(\frac{9}{5} g_1^2 + 3 g_2^2 )- 12 g_5^2 \th(V) \}
\\[10mm]
16 \pi^2 \frac{dY_t}{dt} & = & Y_t \{ 6 Y_t^2 +
Y_b^2 + \frac{3}{2} Y_b^2 \th(H_c) +\frac{3}{2} Y_b^2 \th(H_c')+
\frac{3}{2} Y_t^2 \th(H_c)\nonumber\\[3mm] &+& \frac{3}{2} Y_t^2 \th(H_c')
+Y_{\nu}^2 \th(R)
-(\frac{13}{15} g_1^2 + 3 g_2^2 + \frac{16}{3} g_3^2) -
10 g_5^2 \th(V) \} \\[10mm]
16 \pi^2 \frac{dY_{\nu}}{dt} & = & Y_{\nu}\th(R) \{ Y_{\tau}^2 + 3 Y_t^2 +
2 Y_{\nu}^2 + \frac{3}{2} Y_b^2 \th(H_c) +\frac{3}{2} Y_b^2 \th(H_c')
\nonumber\\[3mm] &+&
2 Y_{\nu}^2 \th(R) + 3 Y_{\nu}^2 \th(H_c) -(\frac{3}{5} g_1^2 +
3 g_2^2) \} \\[10mm]
16 \pi^2 \frac{d\lambda_1}{dt} & = & 3 \lambda_1
\th({\cal{O}}) \{ \frac{4000}{81} \lambda_1^2 + \frac{1152}{29} \lambda_1^2
\th(S) + 144 \lambda_1^2 \th(S_6)+ \frac{128}{3} \lambda_1^2 \th(S_3)
 \nonumber\\[3mm]&+& \frac{160}{3} \lambda_1^2 \th(S_8) + \frac{1744}{27}
\lambda_1^2 \th(V)
- 6 g_3^2 -10 g_5^2 \th(V)\}
\end{eqnarray}
Note that $\th(R)$ stands for $\th(Q^2-M_R^2)$ and that again
we consider cases with $M_V=M_{GUT}$.

We have taken for all figures arranged below the input soft breaking
masses at $M_{GUT}$ to be: $A_o=400 GeV$, $M_o=300 GeV$ and $M_{1/2}=
300 GeV$. Also the superheavy mass $M_{\Sigma}$ is taken to be
$M_{\Sigma}=10^{16} GeV$. These values are indicative values
corresponding to acceptable $\alpha_s$ $\cite{dedes2}$.

In Figure 2 we have plotted the ratio of the $b-\tau$ Yukawa couplings
at $M_{GUT}$ as a function of $tan\beta(M_Z)$ when we vary the
pole mass of the bottom quark in the
range allowed by Table I.
It is clear in the case
of the MDM+PQ model we get $b-\tau$ Yukawa unification both
for $\mu>0$ or $\mu<0${\footnote { We follow the conventions
of Ref.$\cite{dedes1}$}
 together with the excellent agreement of the
value of $\alpha_s(M_Z)$ as this is compared with the experimental
average value $0.118\pm 0.003$.
 In the case of $\mu>0$ for large
$\tan\beta$ there are significant finite threshold corrections
in the bottom and tau Yukawa couplings which arise from the chargino/squark
or gluino/squark
loops and they are proportional to $\mu tan\beta$.
This fact $\cite{carena}$ supports
$b-\tau$ Yukawa unification in the large value region of $tan\beta$ for
$\mu>0$.
 Note also that the upper bound
in the input parameter $tan\beta(M_Z)\leq 36$ does {\it not} come
from the bound of the proton decay rate as in the case of the
minimal $SU(5)$ model, but from the requirement of the radiative
symmetry breaking. This is a special feature of the {\it Peccei-
Quinn} version of the {\it Missing Doublet Model} and is valid for
all figures arranged here.

In Figure 3 we display the ratio of both $b-\tau$ and $t-b$
Yukawa couplings as the top pole mass is varied
inside the region $m_t^{pole}=182\pm 8 GeV$ in the MDM+PQ
model.
The $\frac{Y_b}{Y_{\tau}}$ ratio then takes values in the
regions $0.91-1.09$ for $\mu<0$ and $1.00-1.11$ for
$\mu>0$ respectively.
Note also that as can be seen in Fig.3c and Fig.3d the $t-b$ ratio
never reaches unity due to the
constraint of the radiative symmetry breaking.

Proceeding next to Figure 4 we vary the mass of the colour triplet
$M_{H_c}$(Fig4a,b) and the mass of the intermediate
doublet $M_{H_f'}$ (Fig4c,d) for $\mu<0$ and $\mu>0$ keeping
fixed the $m_t^{pole}=180 GeV$ and $m_b^{pole}=4.9 GeV$.
Acceptable values for $\alpha_s$ are obtainable if we stay in
the vicinity of $M_{H_c}=10^{15} GeV$ and $M_{H_f'}=10^{11} GeV$
$\cite{dedes2}$. For larger values of the colour triplet or
smaller values of the intermediate doublet we get larger values for
the strong coupling $\alpha_s$ completely ruled out by
experiment.
As can be seen in Figure 4 $b-\tau$ unification seems to prefer the
case where $\mu<0$ and low values of $tan\beta\simeq 5$. If we lower
the $m_b^{pole}$ down to $4.8 GeV$ we also
get for $\mu>0$ (this is shown in Fig.2)
$b-\tau$ unification in the low and large $tan\beta$ regime.

In Figure 5 we
consider the behavior of all ratios of Yukawa couplings.
The Renormalization group equations for the three ratios
$R_{t-\nu}\equiv\frac{Y_t}
{Y_{\nu}}$, $R_{b-\tau}\equiv\frac{Y_b}{Y_\tau}$ and
$R_{t-b}\equiv\frac{Y_t}{Y_b}$ are
\begin{eqnarray}
\frac{dR_{t-\nu}}{dt}& =& R_{t-\nu}\{3 (Y_t^2 - Y_{\nu}^2 \th(R))+
(Y_b^2 - Y_{\tau}^2) + 3 (Y_t^2 - Y_{\nu}^2) \th(H_c)\nonumber\\[3mm]
 &-&
(\frac{4}{15} g_1^2 + \frac {16}{3} g_3^2 ) -10 g_5^2 \th(V) \}
\\[5mm]
\frac{dR_{b-\tau}}{dt} &=& R_{b-\tau} \{ 3 (Y_b^2 - Y_{\tau}^2) +
(Y_t^2 - Y_{\nu}^2 \th(R)) + (Y_{\nu}^2 -Y_t^2) \th(H_c)\nonumber\\[3mm]
&-&
(\frac{16}{3}g_3^2 - \frac{4}{3}g_1^2) + 4 g_5^2 \th(V) \}
\\[5mm]
\frac{dR_{t-b}}{dt}&=&R_{t-b} \{ 5 (Y_t^2 - Y_b^2) - (Y_{\tau}^2 -
Y_{\nu}^2 \th(R) ) + (Y_t^2 - Y_{\nu}^2) \th(H_c) \nonumber\\[3mm]
&-&\frac{2}{5}g_1^2 -
2 g_5^2 \th(V) \}
\end{eqnarray}

It can be easily observed that the $SO(10)$-type condition
$Y_t=Y_b=Y_{\tau}=Y_{\nu}$ in the limit of high energies
where all thresholds theta functions are equal to 1
corresponds a zero of the beta functions for the ratios
provided we neglect the contribution of gauge couplings.
In contrast to that the $SU(5)$ condition $Y_b=Y_{\tau}$
leads to a zero of the $\beta_{R_{b-\tau}}=\frac{dR_{b-\tau}}{dt}$
function in the presence of gauge couplings. Note that neither of
the above occurs in MSSM where the contributions of the extra
fields imposed by the extended GUT symmetry are absent.

Figure 5 shows that the $SO(10)$-like Yukawa unification requires as
expected large values of $tan\beta$ which are not allowed by
the requirement of radiative symmetry breaking.
Note however that this is not necessarily true in a
realistic $SO(10)$ model.
Although
complete Yukawa unification is excluded $Y_b=Y_{\tau}$ and
$Y_t=Y_{\nu}$ can be realized independently for a wide range of
$tan\beta$ values. It should be noted that the neutrino pole-mass
that corresponds to $t-\nu$ Yukawa coupling unification is
approximately $m_{{\nu}_{\tau}}^{pole}=(100 GeV)^2/M_R$. This,
very roughly, corresponds to a low energy value of the top
Yukawa coupling
approximately twice that of the neutrino Yukawa coupling.
For a cosmologically interesting neutrino of
$m_{{\nu}_{\tau}}^{pole}=
10 eV$ this fixes the lepton number-PQ-breaking scale to be
$M_R=10^{12} GeV$.

\section{Brief Conclusions}

In this article we have studied in a
2-loop accuracy framework  the unification of the Yukawa couplings
in extensions of the MSSM which are based on the group
$SU(5)$.
Both logarithmic and finite threshold contributions
of SUSY particles have been included in the
calculation of the gauge and Yukawa couplings at $M_Z$.
We have taken into account the effect of all heavy masses
in the renormalization group for the gauge and Yukawa couplings.
The models that we have examined are the {\it minimal $SU(5)$}
model and the {\it Peccei-Quinn} version of the {\it Missing Doublet}
model.
In the minimal $SU(5)$ model $b-\tau$ unification is
possible for $m_b$ values larger than $5 GeV$ in the
low $tan\beta$ regime but in all cases the encountered
$\alpha_s(M_Z)$ values are too large to reconcile with
experiment.
Due to the bound of the proton decay stability we cannot
go to the large $tan\beta$ regime in this model. On the other
hand, in the {\it Peccei-Quinn} version of the {\it Missing Doublet}
model, where  the effect of the right handed neutrino is
naturally included, we have found (Fig.2,3,4)
 that $b-\tau$ unification at $M_{GUT}$ can
be achieved both in the low ($\mu>0$ and $\mu<0$)
and the large ($\mu>0$) $tan\beta$ regime for
$\alpha_s$ values compatible with the low
energy experimental data.
It is interesting to note that this is achieved for
value of the intermediate {\it Peccei-Quinn} $b-\tau$ lepton
number breaking scale in the cosmologically interesting
neighborhood
of $10^{12} GeV$.
 We have also considered whether all
the Yukawa couplings are unified at the scale $M_{GUT}$ (Fig.5). Only
$b-\tau$  Yukawa coupling unification
is possible within this model since the constraint of the radiative
symmetry breaking forbids an $SO(10)$ like condition
$Y_t=Y_b=Y_{\tau}=Y_{\nu}$ at $M_{GUT}$.
\vspace{.5 cm}

{\bf Acknowledgments}
\vspace{.5 cm}

We would like to thank A. B. Lahanas and J. Rizos
for useful discussions and
comments.  Both acknowledge support from the
Program $\Pi$ENE$\Delta$ 95 of the Greek Ministry
of Research and Technology.

\newpage

\par
\vspace{.25in}
\centerline{\psfig{figure=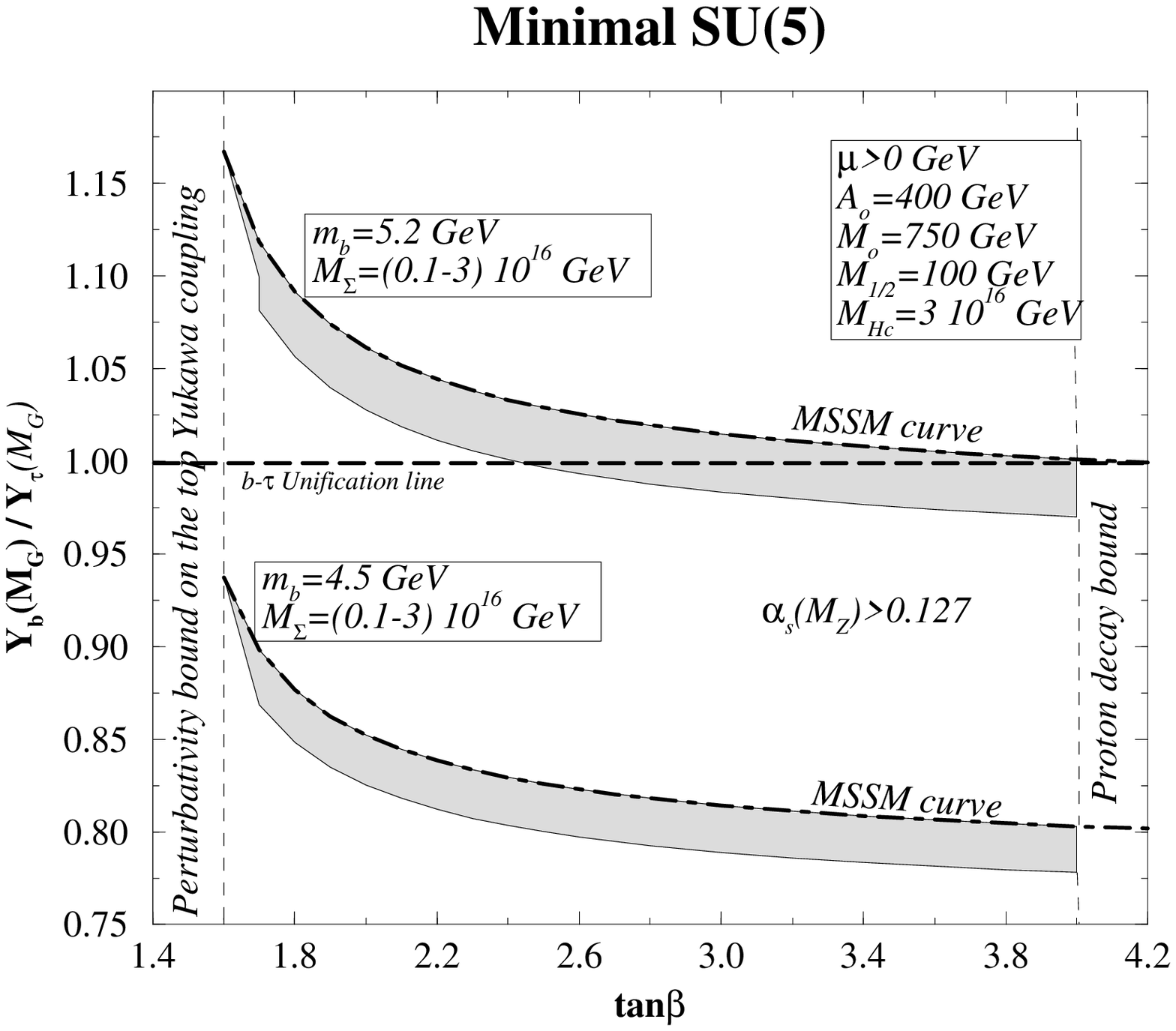}}
{\footnotesize{{\bf Figure 1:} The allowed region for the ratio of bottom
and tau Yukawa couplings at $M_{GUT}$. The input values of the tau lepton
 and top
quark are $m_{\tau}=1.777 GeV$ and $m_t=180 GeV$ respectively.
The extracted value of $\alpha_s(M_Z)$ is greater than .127 . If we still
lower the $M_{\Sigma}$ value down to $10^{14} GeV$, we observe quite
 large values of $\alpha_s$.
In the upper curve of the shaded regions the heavy particles $M_{\Sigma}$,
$M_{H_c}$ have been decoupled (MSSM curve).}}
\par
\vspace{.25in}
\centerline{\psfig{figure=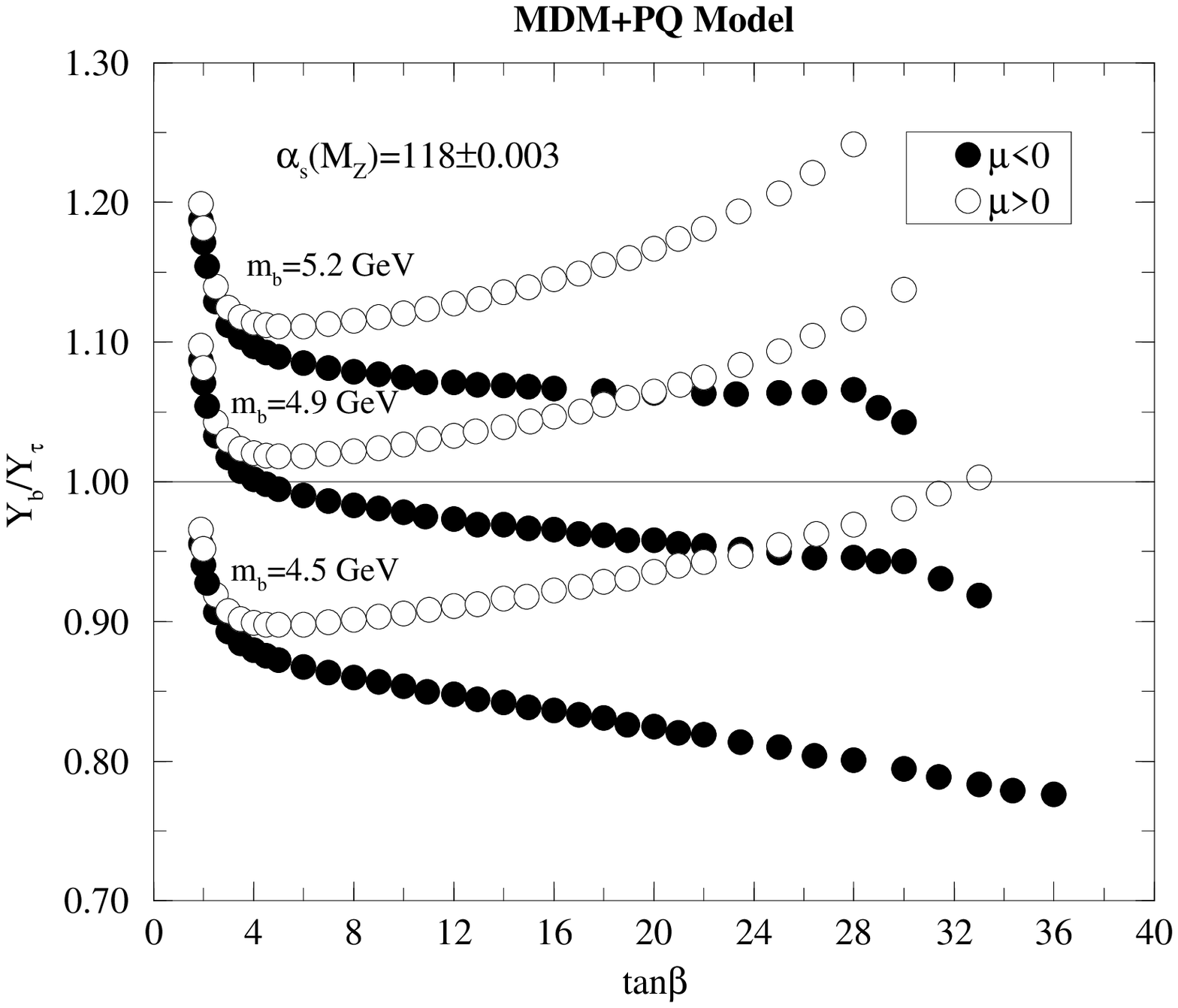}}
{\footnotesize{{\bf Figure 2:} The ratio of the $b-\tau$ Yukawa couplings
as a function of $\tan\beta$  in the MDM+PQ model
when the $m_b^{pole}$ is varied for $\mu<0$ and  $\mu>0$. The
fixed input parameters are $m_t^{pole}=180 GeV$, $M_{H_f'}=10^{11} GeV$,
$M_{H_c}=10^{15} GeV$, $M_{R}=10^{11} GeV$, $m_{{\nu}_{\tau}}^{pole}=
10 eV$.
}}
\par
\vspace{.25in}
\centerline{\psfig{figure=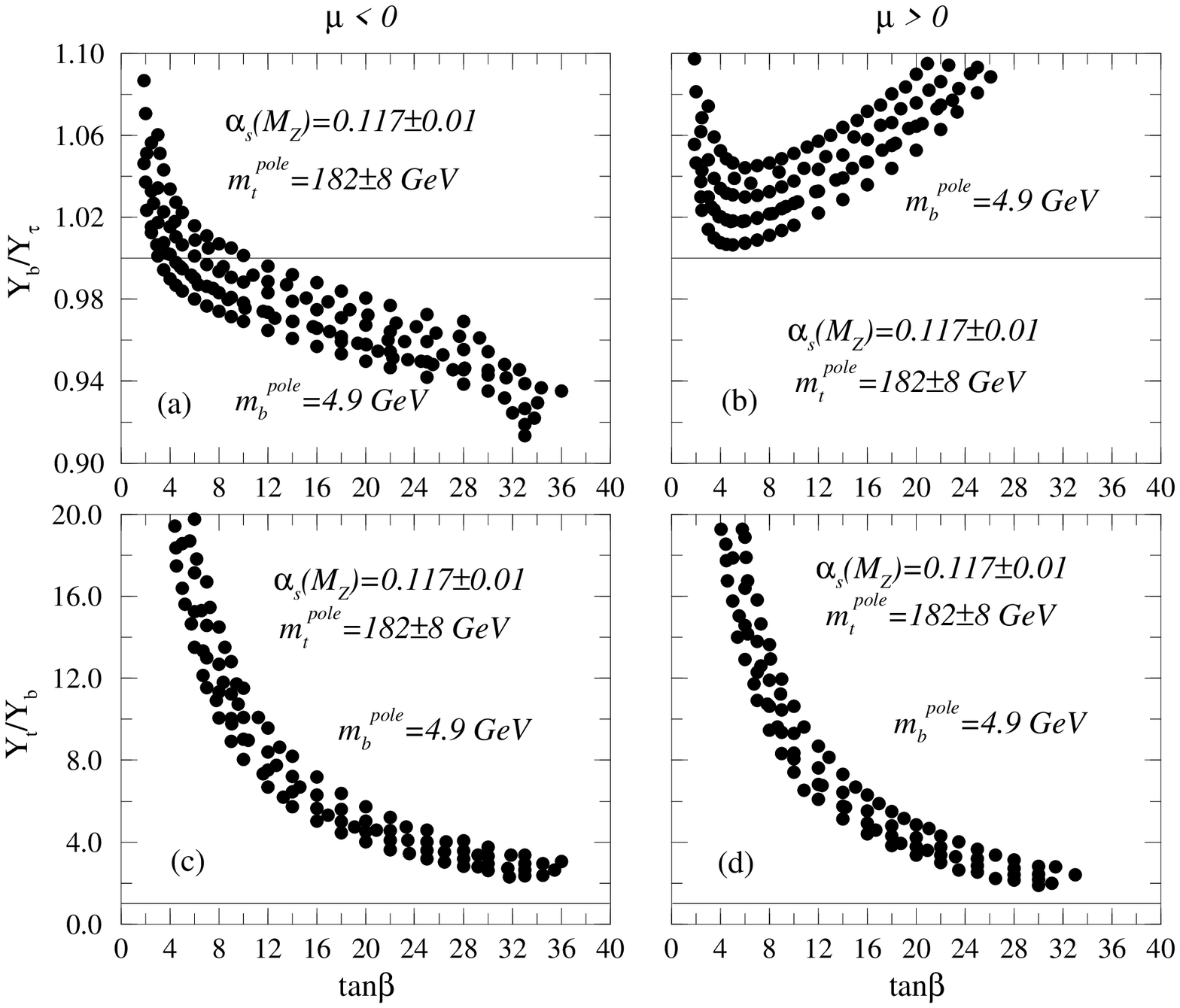}}
{\footnotesize{{\bf Figure 3:} The ratio of the $b-\tau$ and $t-b$ Yukawa
couplings at $M_{GUT}$ as the top pole mass is varied. We keep
the bottom pole mass,  $m_b^{pole}=4.9 GeV$ fixed.
The horizontal line shows the $b-\tau$ or $t-b$ unification.
The other inputs are similar to those displayed in the Figure 2.
}}
\par
\vspace{.25in}
\centerline{\psfig{figure=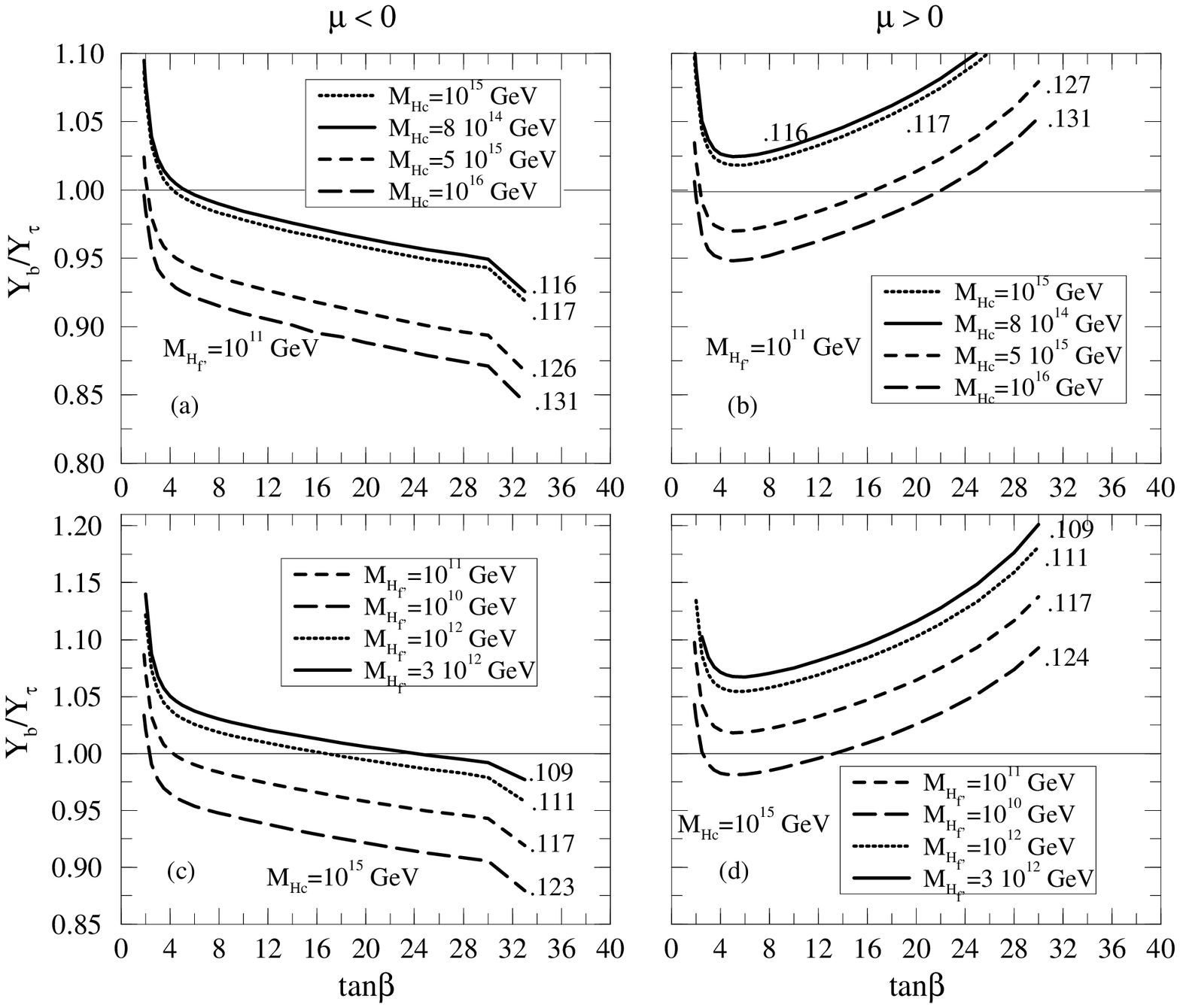}}
{\footnotesize{{\bf Figure 4:} The effect of the variation of the $M_{H_c}$
and $M_{H_f'}$, within the range allowed by proton stability
in the $b-\tau$ unification and also in the extracted
value of $\alpha_s$. We take as inputs: $m_b^{pole}=4.9 GeV$ and
$m_t^{pole}=180 GeV$.
}}
\par
\vspace{.25in}
\centerline{\psfig{figure=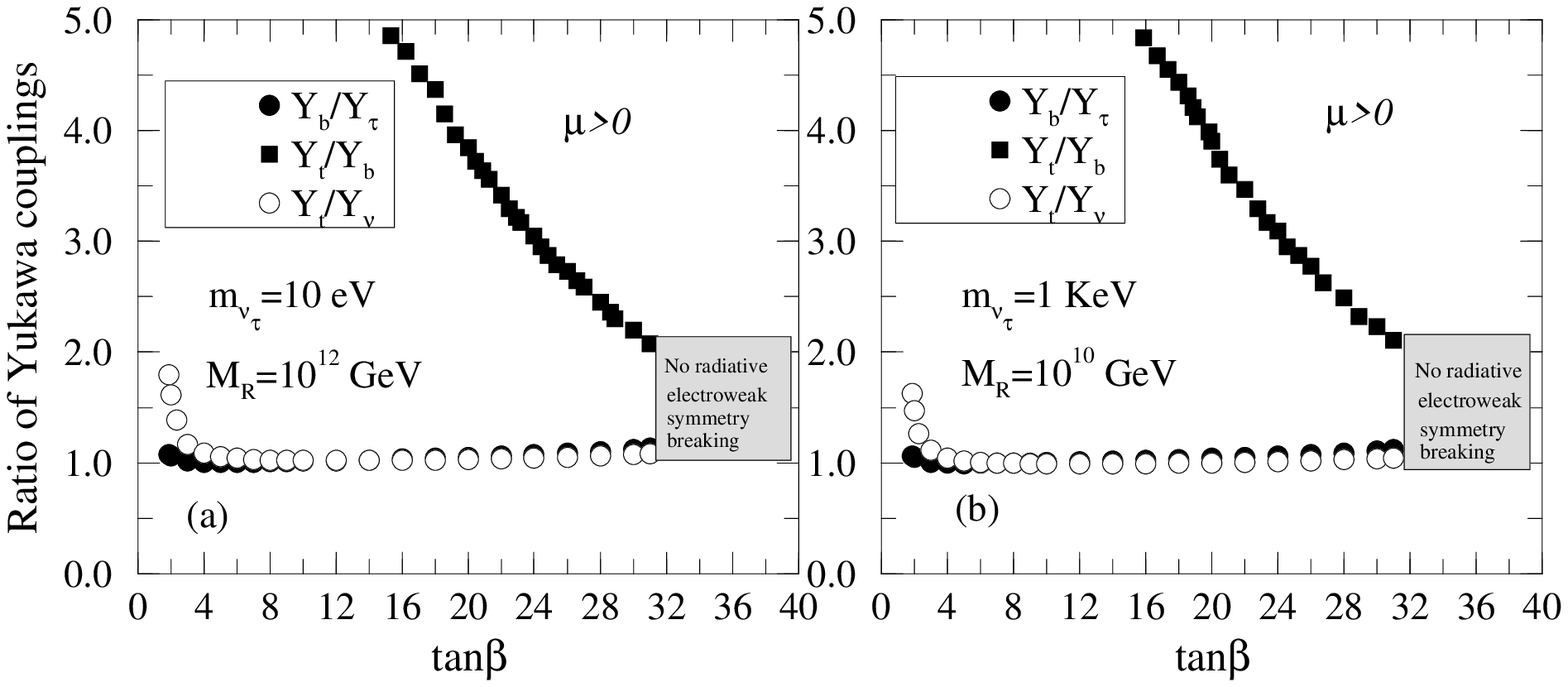}}
{\footnotesize{{\bf Figure 5:} Unification of all Yukawa couplings in
the MDM+PQ model with $m_b^{pole}=4.9 GeV$, $m_t^{pole}=180 GeV$
and the other input parameters displayed in Fig.2
}}


\begin{thebibliography}{99}
\bibitem{Nilles}For reviews see: \\
H. P. Nilles, Phys. Rep. 110(1984)1 ;

H. E. Haber and G. L. Kane, Phys. Rep. 117(1985)75 ;

A. B. Lahanas and D. V. Nanopoulos, Phys. Rep. 145(1987)1.
\bibitem{Iban}  L. E. Iba\~nez  and G. G. Ross, Phys. Lett.
110(1982)215;

K. Inoue, A. Kakuto, H. Komatsu and S. Takeshita, \\
Progr. Theor. Phys. 68(1982)927, 71(1984)96 ;

J. Ellis, D. V. Nanopoulos and K. Tamvakis, Phys. Lett. B121(1983)123;

L. E. Iba\~nez , Nucl. Phys. B218(1983)514 ;

L. Alvarez-Gaum\'e, J. Polchinski and M. Wise, Nucl. Phys. B221(1983)495;

J. Ellis, J.S. Hagelin, D.V. Nanopoulos and K. Tamvakis,\\ Phys. Lett.
B125(1983)275;

L. Alvarez-Gaum\'e, M. Claudson and M. Wise, Nucl. Phys. B207(1982)96;

C. Kounnas, A. B. Lahanas, D. V. Nanopoulos and M. Quiros,\\
Phys. Lett. B132(1983)95 , Nucl. Phys. B236(1984)438;

L. E. Iba\~nez and C. E. Lopez, Phys. Lett. B126(1983)54 ,
Nucl. Phys. B233(1984)511.
\bibitem{Ross}
G. G. Ross and R. G. Roberts, Nucl. Phys. B377(1992)571;

P. Nath and R. Arnowitt , Phys. Lett. B287(1992)89;

S. Kelley, J. Lopez, M. Pois , D. V. Nanopoulos and K. Yuan , \\
Phys. Lett. B273(1991)423 , Nucl. Phys. B398(1993)3;

M. Olechowski and S. Pokorski, Nucl. Phys. B404(1993)590;

M. Carena, S. Pokorski and C. E. Wagner,  Nucl. Phys. B406(1993)59;

P. Chankowski, Phys. Rev. D41(1990)2873;

A.Faraggi, B.Grinstein, Nucl.Phys B422(3)1994.

\bibitem{castano}
G. Gamberini, G. Ridolfi and F. Zwirner, Nucl. Phys. B331(1990)331;

R. Arnowitt and P. Nath, Phys. Rev. D46(1992)3981;

D. J. Casta\~{n}o, E. J. Piard and P. Ramond, Phys. Rev. D49(1994)4882;

G.L Kane, C.Kolda, L.Roszkowski and J.Wells, Phys.Rev D49(6173)1994;

V.Barger, M.S.Berger, P.Ohmann, Phys.Rev.D49(4900)1994.
\bibitem{amaldi}
J. Ellis, S. Kelley and D. V. Nanopoulos, Phys. Lett. B260(1991)131;

U. Amaldi, W. De Boer and M. F\"{u}rstenau,
 Phys. Lett. B260(1991)447;

P. Langacker and M. Luo, Phys. Rev. D44(1991)817.
\bibitem{dedes1}
A. Dedes, A. B. Lahanas and K. Tamvakis, Phys. Rev. D53(1996)3793.
\bibitem{dedes2}
A. Dedes, A. B. Lahanas, J. Rizos and K. Tamvakis, Phys. Rev. D55(1997)2955.
\bibitem{Bagger}
J. Bagger, K. Matchev and D. Pierce, Phys. Lett. B348(1995)443.
\bibitem{Bagger2}
J. Bagger, D. Pierce, K. Matchev and R. Zhang, SLAC-PUB-7180, hep-ph/9606211.
\bibitem{donini}
A. Donini, Nucl. Phys. B467(1996)3.
\bibitem{Slansky}
T. Yanagida, in Proceedings of Workshop on the Unified Theory and the
Baryon Number in the Universe, Tsukuba, Japan,1979, edited by
A. Sawada and A. Sugamoto (KEK,Tsukuba,1979),p.95;

M. Gell-Mann, P. Ramond and R. Slansky, in Supergravity, proceedings
of the workshop, Stony Brook, New York,1979, edited by P. Van
Nieuwenhuizen and D. Z. Freedman (North-Holland, Amsterdam,1979),
p.315.
\bibitem{lah}
A. B. Lahanas and K. Tamvakis, Phys. Lett. B348(1995)451.
\bibitem{booklet}
R. M. Barnett et al., Phys. Rev. D54(1996)1.
\bibitem{pich}
M. Jamin, A. Pich, hep-ph/9702276.
\bibitem{carena}
M. Carena, M. Olechowski, S. Pokorski, C.E.M Wagner, Nucl. Phys. B426(1994)269.
\bibitem{wright}
B.D. Wright, hep-ph/940421;

R. Rattazzi and U. Sarid, Phys. Rev. D53(1996)1553;

N. Polonsky, Phys. Rev. D54(1996)4537.
\bibitem{visani}
F. Vissani and A. Y. Smirnov, Phys. Lett. B341(1994)173;

G. K. Leontaris, S. Lola and G. G. Ross, Nucl. Phys. B454(1995)25.
\bibitem{brig}
A. Brignole, H. Murayama and R. Rattazzi, Phys. Lett. B335(1994)345;
\bibitem{dimopoulos}
S. Dimopoulos, H. Georgi, Nucl. Phys. B193(1981)150;

N. Sakai, Z. Phys. C11(1981)153.
\bibitem{bachas}
C. Bachas, C. Fabre and T. Yanagida, Phys. Lett. B370(1996)49.
\bibitem{masiero}A. Masiero, D.V. Nanopoulos, K. Tamvakis and T. Yanagida,
Phys.Lett. B115(1982)380;

B. Grinstein, Nucl. Phys. B206(1982).
\bibitem{peccei}
R. Peccei and H. Quinn, Phys. Rev. Lett. 38(1977)1440;

R. Peccei and H. Quinn, Phys. Rev. D16(1977)1791.
\bibitem{hisano}J. Hisano, T. Moroi, K. Tobe, and T. Yanagida, Phys. Lett.
B342(1995)138;

J. Hisano, TIT-HEP-307, Talk given at Yukawa International Seminar'95;

J.L Lopez, D.V. Nanopoulos, Phys. Rev. D53(1996)2670.
\bibitem{murayama}
H. Murayama, H. Suzuki and T. Yanagida, Phys. Lett. B291(1992)418.

\end{thebibliography}
\end{document}